\documentclass{aa}
\usepackage{graphicx}
\usepackage{txfonts}
\usepackage[hypertexnames=false]{hyperref}
\hypersetup{colorlinks=true, citecolor=blue}
\usepackage{algorithm}
\usepackage{algpseudocode}

\renewcommand{\linenumbers}{}

\begin{document}

    \title{Interpreting anomaly detection of SDSS spectra}

    \author{E. Ortiz \inst{1} \and M. Boquien \inst{2}}

    \institute{
        Departamento de F\'isica, Universidad de Antofagasta
        Avenida Angamos 601, Antofagasta, Chile.
        \email{edgar.ortiz@ua.cl}
        \and
        Université Côte d'Azur, Observatoire de la Côte d'Azur, CNRS, Laboratoire Lagrange, 06000, Nice, France
        \email{mederic.boquien@oca.eu}
    }

    \date{Received -- --, ----; accepted -- --, ----}

    \abstract
        { 
            The increasing use of machine learning in astronomy introduces important questions about interpretability.
            Due to their complexity and non-linear nature, it can be challenging to understand their decision-making process, especially when applied to anomaly detection. 
            While these models can effectively identify unusual spectra, interpreting the physical nature of the flagged outliers remains a major challenge.
        }
        {
            We aim to bridge the gap between anomaly detection and physical understanding by combining deep learning with interpretable machine learning (iML) techniques to identify and explain anomalous galaxy spectra from SDSS data.
        }
        {
            We present a flexible framework that uses a variational autoencoder to compute multiple anomaly scores, including physically-motivated variants of the mean squared error. We adapt the iML LIME algorithm to spectroscopic data, systematically explore segmentation and perturbation strategies, and compute explanation weights that identify the features most responsible for each detection. To uncover population-level trends, we normalize the LIME weights and apply clustering to the top 1\% most anomalous spectra.
        }
        {
            Our approach successfully separates instrumental artifacts from physically meaningful outliers and groups anomalous spectra into astrophysically coherent categories. These include dusty, metal-rich starbursts; chemically-enriched H\,II regions with moderate excitation; and extreme emission-line galaxies with low metallicity and hard ionizing spectra. The explanation weights align with established emission-line diagnostics, enabling a physically-grounded taxonomy of spectroscopic anomalies.
        }
        {
            Our work shows that interpretable anomaly detection provides a scalable, transparent, and physically meaningful approach to exploring large spectroscopic datasets. Our framework opens the door for incorporating interpretability tools into quality control, follow-up targeting, and discovery pipelines in current and future surveys.
        }


    \maketitle

\section{Introduction}

    Thanks to recent advances in instrumentation, and telescope technology \citep{moravec2019, siemiginowska2019}, astronomy is entering in an era of abundance. This abundance poses new challenges. The traditional knowledge discovery process does not scale well in this scenario of big and complex data. For instance, the Vera Rubin Legacy Survey of Space and Time (LSST) is estimated to deliver about 500 Petabytes worth of data \citep{ivezic2019-lsst}. In this context, the use of Machine Learning (ML) is gaining momentum.

    Among the many types of ML algorithms, Deep Neural Networks (DNNs) are becoming ubiquitous. While their flexibility and the affinity of these models with GPUs has made them prime choices for addressing many challenges, employing them often comes at the expense of interpretability: the non-linear interactions between the features of the data and the different parameters of DNNs make the problem intractable in most cases. 
    This opacity, often referred to as the `black box' problem, is a widely recognized challenge in applying advanced Artificial Intelligence (AI) to scientific domains, as it can hinder trust, verification, and reproducibility \citep{adadi2020-peeking-inside-black-box, molnar2020, ema2021, 202502_li_machine_learning_stellar_astronomy, 202503_wetzel_interpretable_machine_learning_physics, 202506_lieu_interpretable_ai_astronomy}. 
    In this sense, DNNs conform black boxes.

    The field of interpretable Machine Learning (iML), or Explainable AI (XAI), aims to demystify these complex models, providing human-friendly explanations for a model's predictions. Given the trend of ML adoption in astronomy, iML techniques are becoming a need for ensuring that AI-driven research remains reliable, transparent, and aligned with scientific principles. The use of iML in science facilitates the debugging and improvement of models, builds trust by verifying that predictions are based on sound and well established physics and phenomenology, and, most importantly, accelerates knowledge acquisition by providing insights into the underlying processes a model has learned from data, potentially leading to new hypotheses and discoveries \citep{adadi2020-peeking-inside-black-box, 202412_sahakyan_ai_cosmos, 202503_wetzel_interpretable_machine_learning_physics}.

    Recent years have seen a growing interest in applying iML techniques in astronomy, with use cases spanning galaxy morphology, spectroscopic classification, photometric redshift estimation, and cosmology. These methods fall into two main categories. The first includes model-specific explanations, which are intrinsically tied to the architecture or training process of the model itself \citep{
    202111_yip_peeking_inside_black_box_sensitivity_analysis,
    202201_jacobs_iml_gravitational_lens_sensitivity,
    202204_bhambra_galaxy_morphology_saliency_maps,
    202209_lucie_insights_halo_mass_profiles_ml_intrinsically_interpretable, 
    202212_gully_interpretable_ml_spectroscopy_intrinsic_interpretability, 
    202310_pandey_explainable_solar_flares_intrinsic_plus_shap, 
    202504_bonse_use_the_4s_signal_safe_speckle_subtraction_saliency_maps}.
    The second category comprises model-agnostic post-hoc methods, such as Local Interpretable Model-agnostic Explanations \citep[LIME,][]{ribeiro2016} and SHapley Additive exPlanations \citep[SHAP,][]{201705_lundberg_shap_paper, lundberg2020}, which can be applied to any predictive model. Within this class, SHAP-based studies have been used to understand model behavior across applications including SED modeling, galaxy clustering, molecular abundance prediction, and cosmological parameter inference \citep{
    202107_gilda_sed_modeling_shap,
    202110_poletti_SHAPing_the_gas, 
    202204_villaescusa_cosmology_one_galaxy_shap,
    202210_dey_photometric_redshifts_sdss_capsule_network_shap, 
    202309_heyl_statistical_machine_learning_astrochemistry_shap,
    202311_heyl_understanding_molecular_abundances_shap, 
    202403_crupi_enhancing_gamma_ray_burst_detection_shap, 
    202412_elvitigala_galaxy_clustering_shap_lime_more, 
    202502_grassi_mapping_synthetic_observations_prestellar_core_models_shap, 
    202505_ye_deep_iml_analysis_carbon_star_gaia_dr3_shap}.
    On the other hand, LIME-based approaches have been used in studies of strong emission-line galaxies, black hole accretion, and cosmological model selection \citep{
    202207_dold_galaxy_clusters_idnetification_lime, 
    202404_pasquato_iml_black_holes_lime_anchors, 
    202501_ocampo_iml_cosmological_model_selection_lime}.
    Together, these efforts show the growing role of iML in astronomy and its potential to improve trust, transparency, and scientific interpretability.

    Anomaly detection represents a particularly compelling application for iML. As data volumes keep growing, the ability to automatically flag outliers in surveys like the Sloan Digital Sky Survey (SDSS), Gaia, or LSST becomes critical for identifying rare or unexpected phenomena. The field has advanced significantly with the development of unsupervised and self-supervised ML methods that detect outliers in images, spectra, and time-series data \citep[e.g.,][]{
    baron2017,
    reis2018-outliers-apogee,
    ichinohe2019,
    giles2020,
    lochner2021,
    202301_vafaei_sadr_personalized_anomaly_detection,
    202306_ciprijanovic_deepastrouda_cross_survey_gal_morph_anomaly,
    202403_etsebeth_astronomaly_scale_searching_anomalies,
    202410_aleo_anomaly_detection_transients_real_time,
    202503_lochner_astronomaly_protege_human_machine_collaboration,
    202504_semenikhin_real_bogus_scores_active_anomaly_detection,
    202507_kornilov_coniferest_active_anomaly_detection_framework}.
    Yet, detection alone is insufficient for scientific insight. Many flagged outliers remain poorly characterized, and the sheer volume of candidates overwhelms manual inspection efforts. In this context, iML offers a promising solution by helping to diagnose why an object is considered anomalous, highlighting which features contribute most to the anomaly score. This not only improves trust and interpretability, but also enables astronomers to prioritize the most compelling candidates.

    In this work, we present a new and flexible public tool to easily perform explainable anomaly detection on spectroscopic observations based on LIME. In Sect.~\ref{section: data} we present the SDSS Data Release 16 (DR16) spectroscopic sample we are using and the data cleaning procedures we perform in preparation for the anomaly detection algorithm that we introduce in Sect.~\ref{section: anomaly}. We detail the LIME algorithm adaptation for interpretable anomaly detection in Sect.~\ref{section: iml} before presenting the results in Sect.~\ref{section: results} and discussing them in Sect.~\ref{section: discussion} and concluding in Sect.~\ref{section: conclusions}.

\section{Sample selection and data processing} \label{section: data}

    For this work, we adopt the SDSS DR16 spectroscopic dataset of galaxies. Its reasonably large size and excellent quality, make it ideal for anomaly detection as shown in \cite{baron2017}.

    The data and associated metadata were obtained from the skyserver data repository\footnote{\url{http://skyserver.sdss.org/CasJobs/}}, yielding a total of 1,230,784 spectra from the ``sdss'' dataset. We shift the spectra to their rest frame and interpolate them to a common grid in the optical region of $350-750$~nm. High-redshift spectra have many missing values. To avoid such scenario we constrain our sample to be in the redshift range between $0.01-0.50$. As a consequence we are left with a total of 791,738 spectra. To avoid contamination by the [OI$\lambda$557.7] line, we remove the wavelengths in the region of $556.5-559.0$~nm from all spectra before de-redshifting. Afterwards, we remove wavelengths with a signal-to-noise smaller than 1. We also correct for the Milky Way foreground extinction, according to dust maps from \cite{schlegel1998-ebv}\footnote{\url{https://github.com/kbarbary/sfddata}}. Once the spectra have been corrected for the foreground extinction, we shift them to rest frame and all the spectra are interpolated to a common grid with a resolution of 0.1~nm. After the interpolation we drop all the spectra with more than 10\% of NaN values leaving us with a final sample of 728,133 spectra. The summary statistics for redshift and median SNR of this sample are presented in Table \ref{table: final-sample}.
    \begin{table}[!ht]
        \caption{
            Summary statistics for redshift (z) and median SNR of final sample.
        }
        \label{table: final-sample}
        \centering
        \begin{tabular}{ccc}
            \hline\hline
            Statistics  &    z    &  Median SNR \\
            \hline
                mean    &    0.11 &    14.68 \\
                std     &    0.06 &     7.04 \\
                min     &    0.01 &     4.00 \\
                25\%    &    0.07 &     9.94 \\
                50\%    &    0.10 &    13.15 \\
                75\%    &    0.15 &    17.55 \\
                max     &    0.30 &    98.10 \\
            \hline
        \end{tabular}
        \tablefoot{
            The final sample contains 728,133 spectra.
            The abbreviations std stands for the standard deviation.
            The min, max abbreviations and the percentages stand for the quartiles of each variable.
        }
    \end{table}

    To further refine the sample, we first drop all wavelengths for which more than 10\% of spectra have Not-a-Number (NaN) values. This reduces the number of wavelength points per spectrum from 4000 to 3773. For the remaining missing values, since ML algorithms are sensitive to missing values and the dynamical range of the data, we replace NaNs in each individual spectrum with its median flux, ensuring consistency. We then apply a median-flux normalization to each spectrum to bring them onto a common scale, mitigating the effect of absolute flux variations. Finally, considering that machine learning algorithms are sensitive to noise levels, we divide the sample into four bins based on the median SNR of each spectrum. This binning helps homogenize noise properties and prevents the anomaly detection model from identifying different noise levels as anomalous features. The first three bins contain 182,094 spectra each, and the fourth bin (highest SNRs) contains 181,850 spectra. For model development, we use 80\% of the spectra in each bin for training and the remaining 20\% for validation (see Sect.~\ref{section: vae}).

\section{Anomaly detection} \label{section: anomaly}

    Anomaly detection is a complex task, in particular with high dimensional data, as is the case of spectra with thousands of fluxes as features.
    In order to perform anomaly detection, we need a measure of the similarity between different observations.
    An object is anomalous if it is dissimilar from the rest; in a scientific context, the goal is to identify which of these dissimilar objects are astrophysically `interesting', a subjective task.
    In high dimensional spaces, the notion of distance is distorted.
    For instance, \cite{aggarwal2001DistanceHighDimension} showed that as the dimensionality of a space increases, the Euclidean metric is not able to discern between the farthest and the closest point to the origin of a data set where all points are generated with a multidimensional uniform distribution.
    One implication of this result is that there is not a single universal anomaly score that is capable of retrieving all types of anomalies, particularly considering that the definition of a `useful' anomaly depends on the scientific goal.
    Similarly, the ``no free lunch theorem'' from optimization theory states that there is not best model for performing a particular task \citep{wolpert1997-no-free-lunch}, a conclusion that directly applies to anomaly detection, for which there is no universal algorithm.
    Different algorithms will retrieve different types of anomalies \citep{nun2016, reis_2021-unsupervised}.

    In this work we implement an anomaly detection framework based on Variational Autoencoders (VAEs), a type of NNs.
    VAEs can be seen as a black box that aims to reconstruct the most common features of spectra.
    As a consequence, spectra with anomalous features will be poorly reconstructed \citep{ichinohe2019}.
    One way to assign an anomaly score is to measure the distance between a spectrum and its reconstruction by the VAE.
    When we couple the VAE with a distance metric, we obtain a regression model that maps a spectrum to an anomaly score.
    We consider different distance metrics to mitigate the curse of dimensionality and be able to retrieve different types of anomalous spectra.
    While some authors identify anomalies as points in low-density regions of the VAE's latent space \citep{portillo2020}, we focus on reconstruction error.
    This approach has the advantage of directly linking an anomaly score to specific misreconstructed features in the spectrum, which aligns naturally with our goal of generating feature-based interpretations.
    In what follows we present a description of VAEs and the distances we use to assign an anomaly score to a spectrum.

\subsection{Variational Autoencoders (VAEs)} \label{section: vae}

    VAEs are neural networks that belong to the family of Generative Learning (GL) models in ML \citep{kingma2013}.
    GL models are designed to generate synthetic data \citep{goodfellow2016book, foster2019GLBook}.
    An example of the use of VAEs with spectra can be found in \cite{portillo2020}, where they are used as a dimensionality reduction technique that outperforms methods such as Principal Component Analysis and Non-negative Matrix Factorization.
    In addition, they show how the lower dimensional representation obtained with VAEs can be used for other downstream tasks, such as generating synthetic data or anomaly detection.
    A VAE is composed of an encoder network and a decoder network.
    The encoder outputs a lower dimensional vector representation of the input spectrum.
    This representation is also called the latent representation.
    The decoder takes the latent representation to produce a reconstruction of the original spectrum.
    Another possibility is to sample points from the latent space, and generate new spectra with the decoder.
    VAEs, by their nature, also allow us to explore perturbations of the data used in the training phase \citep{engel2017}.
    We describe in detail the architecture and the training of the VAE in Appendix \ref{appendix: VAE}.
    While a detailed quantitative comparison with other anomaly detection algorithms is beyond the scope of this paper, it is worth noting that VAEs have proven to be a robust method for this task in astronomy
    \citep[e.g.,][]{ichinohe2019, portillo2020, sanchez_2021_anomaly_agn_vae, liang_2023_anomaly_desi_vae,
    rogers_2024_vae_serendipity_lsst, quispe_2025_autoencoders_sed_anomaly, nicolau_2025_identifying_anomalous_galaxy_spectra_vae}.
    Our VAE performs well in its primary function: it successfully learns to reconstruct the spectra of typical galaxies from the SDSS dataset with low error, as exemplified in the top panel of Figure~\ref{fig: good-bad reconstructions}.
    Consequently, it produces poor reconstructions with high error for spectra containing features not well-represented in the training data (Fig.~\ref{fig: good-bad reconstructions}, bottom panel).
    This behavior demonstrates that the VAE has effectively learned the underlying distribution of the data, making its reconstruction error a reliable proxy for anomaly scoring.

\subsection{Anomaly score}\label{section: score}

    To obtain the anomaly score of a spectrum using the VAE, we follow Algorithm \ref{alg: ODA-score}.
    \begin{algorithm}[!h]
        \begin{algorithmic}
            \Require $X$
                \Comment{$X$: observed spectrum}
            \Require VAE
            \Require S
                \Comment{S: anomaly scoring function}
            \\
            \State $X' \gets$ VAE.reconstruction($X$)
                \Comment{$X'$: reconstruction of $X$}
            \State s $\gets$ S$(X,X')$
                \Comment{$s$: outlier score of $X$}
            \Return score
        \end{algorithmic}
        \caption{Outlier Detection Algorithm}
        \label{alg: ODA-score}
    \end{algorithm}

\begin{figure}[!ht]
        \centering
        \includegraphics[width=.98\columnwidth]
            {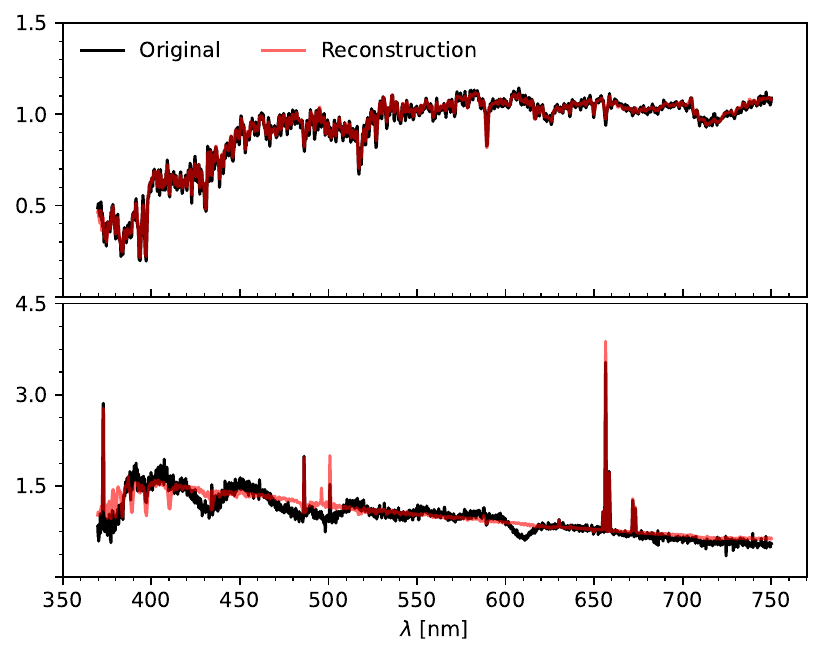}
        \caption{Example of a good (upper panel) and poor (lower panel) reconstruction by the VAE. The MSE score for the good reconstruction is $\approx$0.00019 and for the poor reconstruction is $\approx$0.015. Fluxes are median normalized.}
        \label{fig: good-bad reconstructions}
    \end{figure}

    First we propagate the spectrum through the VAE and obtain its reconstruction. To compute the anomaly score, we use the spectrum, $X$, and its reconstruction, $X'$. By nature, the most common features in the spectra will be well reconstructed and anomalous features will be poorly reconstructed (see Fig.~\ref{fig: good-bad reconstructions} for examples of reconstructions with small and large anomaly scores).

    Then, the key ingredient is the metric to compute the distance, which under Algorithm \ref{alg: ODA-score} is the anomaly score S$(X,X')$. We explore eight anomaly scores, each derived by comparing a spectrum $X$ with its reconstruction $X'$. These scores fall into two families depending on the underlying distance metric: the standard mean squared error (MSE) and an inverse-flux weighted variant ($\chi^2$). The key idea is that poorly-reconstructed features correspond to anomalous content, and each variation is designed to probe a different type of deviation.

    The first family is based on the standard MSE metric (see Eq.~\ref{eq: mse}). We add three physically motivated variations to the MSE to probe different behaviors in spectra. For instance, we notice that the MSE has the tendency to highlight as anomalous, spectra with strong standard emission lines and data glitches. Strong emission lines are anomalous in the sense that they are not very common in our data, but they are objects that in most cases are known, such as startburst galaxies. On the other hand, data glitches are often due to either cosmic rays or instrumental errors. These spectra are anomalous in the same sense that spectra with strong emission lines, but they do not constitute interesting anomalies. The variations we introduce to mitigate the prevalence of these anomalous patterns are:
    \begin{enumerate}
        \item Filtered MSE: masks common emission lines before computing the residuals. This variation mitigates the dominance of strong emission-lines. The filtered MSE is defined as
            \begin{equation}
                MSE_{fnl}(X, X') =
                    \frac{1}{N}
                    \displaystyle \sum_i^N
                    f_i \left(x_i - x_i'\right)^2,
                \label{eq: fnl-mse}
            \end{equation}
        where $f_i$ corresponds to the filter, and takes the value of 1 for fluxes at wavelengths outside the filtered region, and 0 otherwise. The width of the filter is defined by $\Delta \lambda=\lambda_l \Delta v/c$, with $\lambda_l$ the wavelength of the line, $c$ is the speed of light, and $\Delta v$ the velocity width of the masking window. The filtered lines are given in Table~\ref{table: narrow-lines-filter}.
        \item Trimmed MSE: removes the top 3\% largest residuals prior to computing the MSE. This helps suppress the influence of localized glitches such as cosmic rays or bad pixels.
        \item Filtered + Trimmed MSE: applies both line masking and trimming. This combination targets subtle continuum-level anomalies while avoiding common strong features and data artifacts.
    \end{enumerate}
    Together, with the unmodified MSE, these yield 4 total MSE-based scores. The second family uses an inverse-flux weighted MSE, effectively computing the $\chi^2$ between the observation and the reconstruction,
    \begin{equation}
        \label{eq: chiSquare}
        \chi^2(X, X') =
        \frac{1}{N}
        \sum_i^N
        \frac{\left(x_i - x_i'\right)^2}{x_i + \delta},
    \end{equation}
    where $\delta$ is a positive and small value. We introduce $\delta$ to avoid division by zero. This score naturally downweights residuals in bright regions (e.g., strong emission lines), making it complementary to standard MSE. As with MSE, we define three additional variants: (1) Filtered $\chi^2$, (2) Trimmed $\chi^2$, and (3) Filtered + Trimmed $\chi^2$. Together with the baseline $\chi^2$ score, this yields 4 additional $\chi^2$-based scores, for a total of 8 anomaly scoring functions used in our analysis.

\section{Interpretable Machine Learning with LIME} \label{section: iml}

    Understanding what drives the anomaly score for an object is essential for scientific validation, knowledge discovery, and error diagnosis.
    Interpretable Machine Learning provides tools and methodologies to verify that ML-driven conclusions align with physical expectations.
    Such methods can be broadly categorized into model-specific and model-agnostic techniques.
    Model-specific approaches, such as decision trees and integrated gradients in neural networks \citep{201703_sundararajan_integrated_gradients}, have inherent interpretability.
    In contrast, model-agnostic approaches, such as SHAP (SHapley Additive Explanations) \citep{201705_lundberg_shap_paper} and LIME (Local Interpretable Model-agnostic Explanations) \citep{ribeiro2016}, can be applied to any ML model without modifying its internal structure.
    In this work, we introduce LIME-Spectra-Interpreter, an adaptation of LIME for interpreting anomaly detection in spectroscopic data.
    We show how, in the case of anomaly detection, the intuition we gain with LIME-Spectra-Interpreter can help us automate insight by highlighting what features are correlated to the anomaly score of an object.
    To explain a prediction, LIME approximates the complex model with a surrogate that is easier to interpret.
    In this work, we consider the case of linear surrogates.
    An interesting point about LIME is that the space of features of the interpreter is different from the space of features of the model we want to inspect.

    LIME is traditionally used for image and tabular data by perturbing input features (pixels or table columns) and training a simple surrogate model to approximate the complex ML model locally. In image analysis, LIME groups neighboring pixels into superpixels, then selectively removes information to determine feature importance. Our adaptation, LIME-Spectra-Interpreter, extends this idea to spectroscopic data by using spectral segments. Instead of grouping neighboring pixels in an image, we segment spectra into wavelength bins, which can be defined either uniformly or using clustering techniques such as SLIC \citep{SLIC2010}.
    For instance, the interpreter uses segments of the spectrum as predictors, rather than a single wavelength.
    These segments constitute what \cite{ribeiro2016, molnar2020} and \cite{ema2021} denominate human-friendly or meaningful representations of the data. The perturbation process then modifies spectral regions (e.g., by adjusting flux values or adding noise) to examine how these changes impact anomaly scores. 

    To explain the model, LIME perturbs the original data to obtain a new set of spectra.
    These new spectra are then used to fit the interpreter that approximates the original model.
    The process to obtain perturbed spectra makes use of the segmented representation of the spectrum.
    To generate a new spectrum, LIME randomly selects a subset of segments and grays them out.
    All perturbed wavelengths can be set to either a hard coded value defined by the user, or the they can be set to the mean value of their respective segment.

    In mathematical terms, we can see the segmented representation of a perturbed spectrum $Z$ as a binary vector $Z'$, where each entry labels a segment.
    A 1 (respectively 0) value in the $i^{th}$ entry of $Z'$ means that the $i^{th}$ segment of $Z$ is not perturbed (resp. perturbed).
    If we denote the interpreter by $g$, then the interpreter has the form:
    \begin{equation}
        g\left(Z'\right) = \sum_{i=1}^{N} w_i \times z'_i,
        \label{eq: linear-surrogate}
    \end{equation}
    where $N$ is the number of segments and $z'_i$ is the $i^{th}$ entry of $Z'$.
    The weights of the linear model indicate the importance of each segment.
    Because of the random nature of the perturbation process, some spectra are very dissimilar from the original, which might harm the ability of the interpreter to locally approximate the model in the vicinity of the original spectrum. To mitigate this effect, LIME weights the proximity of a perturbation to the original spectrum using an exponential kernel given by:
    \begin{equation}
        \pi_x(z) = \exp \left(
                \frac{-D(x, z)^2}{\sigma^2}
            \right),
        \label{eq: similarity-kernel}
    \end{equation}
    where $D(x, z)$ is the distance between the original spectrum and the perturbation $z$. By default, LIME uses the $L^2$ norm.

    Finally, if we call our model $f$, then the weights of the interpreter are found by minimizing the loss function:
    \begin{equation}
        \mathcal{L}(f, g, \pi_x) =
            \sum_{z, z' \in \mathcal{Z}} \pi_x(z) \left(f(z)-g(z')\right)^2, 
        \label{eq: faithfull-loss}
    \end{equation}
    where $\mathcal{Z}$ is the set of perturbed spectra and their segments representation.

    In summary, our adaptation of LIME for spectroscopic data involves two key modifications. First, instead of image superpixels, we segment spectra into meaningful wavelength regions, using either uniform divisions or clustering-based methods. Second, rather than masking pixels, we perturb the data by scaling fluxes, replacing segments with their mean flux or fixed values to probe model sensitivity to specific features. Throughout this work, we adopt the following default configuration: flux scaling as the perturbation method with a scale factor of 0.9, uniform segmentation (each section indicates number of segments) and 5000 perturbed samples. These settings balance interpretability, resolution, and computational efficiency (see Sect.~\ref{section: explanation-config}).

\section{Results} \label{section: results}
\subsection{Overview of Anomaly Score Findings} \label{results: overview}

\begin{figure*}[!ht]
        \centering
        \includegraphics[width=0.95\textwidth]{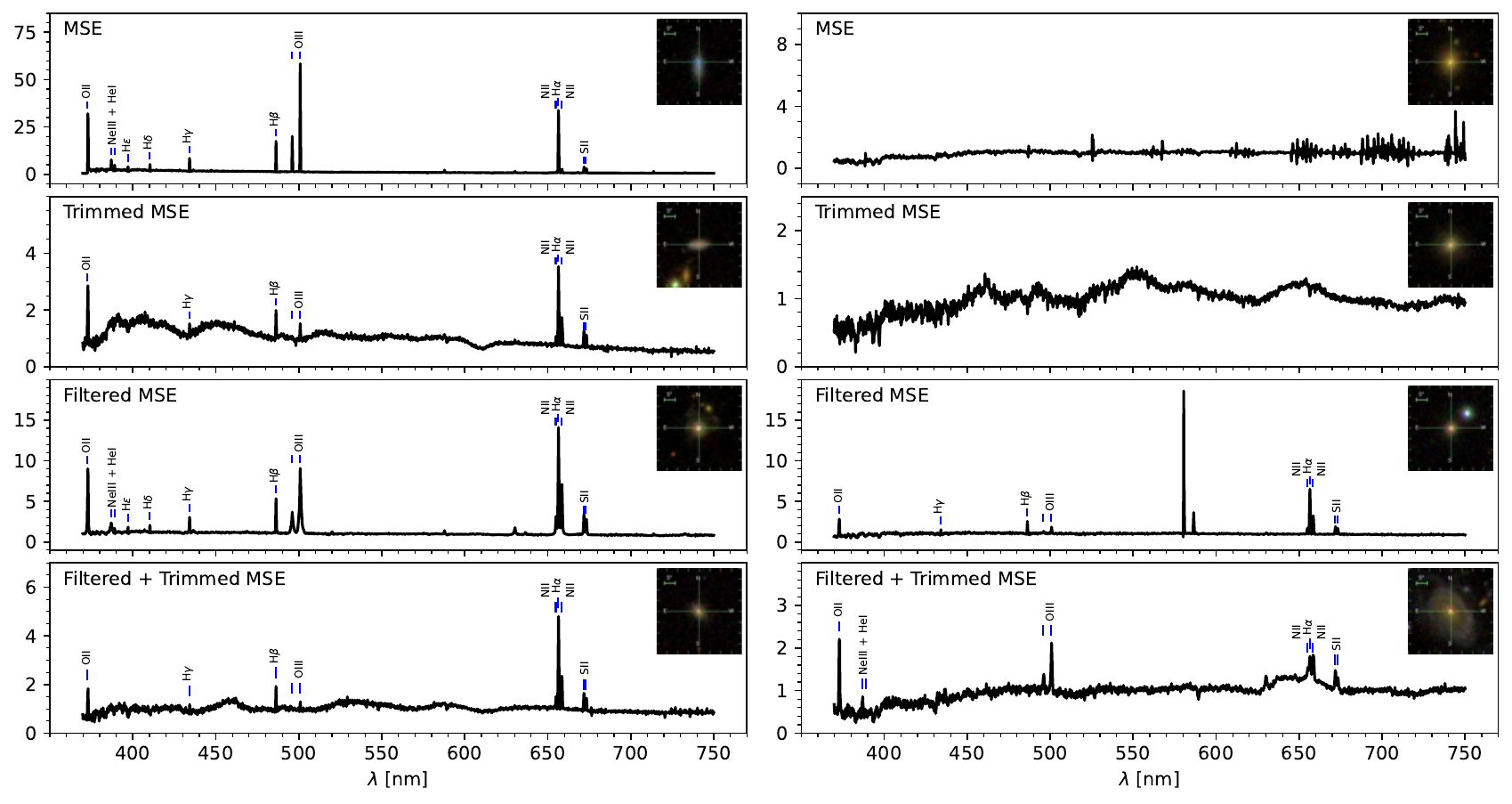}
        \caption{
            Overview of SDSS anomalies identified by different variations of the MSE score.
            Each row displays example spectra (with SDSS imaging thumbnails) found using the respective MSE variation (e.g., MSE, Trimmed MSE), with two representative examples per variation.
            The different MSE variations are effective in highlighting different anomalous patterns in the spectra.
            Fluxes are median normalized.
        }
    \label{anomalies: mse-overview}
    \end{figure*}

    Our anomaly detection framework computes eight anomaly scores: four derived from the MSE and four from inverse-flux weighting ($\chi^2$), each incorporating optional filtering and trimming strategies (Sect.~\ref{section: score}) for each SDSS spectrum.
    Through the inspection of the top anomalous spectra we find that different scoring metrics emphasize distinct spectral characteristics, revealing a variety of astrophysical and instrumental anomalies.
    We present a sample of anomalous spectra in Fig.~\ref{anomalies: mse-overview}.

    We observe that the MSE metric (top row) primarily identifies spectra with strong emission-line features, particularly in the [OIII] and H$\alpha$ regions. This suggests a sensitivity to extreme star-forming galaxies where these emission lines dominate the flux distribution. Additionally, we find MSE detects spectra with data artefacts.

    When the MSE score is computed after ignoring 3\% of the largest residuals (second row), its focus shifts toward anomalies with unusual continua. This modification reduces sensitivity to extreme flux variations and instead highlights objects with peculiar stellar populations or spectra that do not correspond to a typical galaxy.

    Applying a 250~km~s$^{-1}$ filter to the locations of standard emission lines (third row) further refines the selection, prioritizing spectra with broader emission lines. This variation detects anomalies likely corresponding to turbulent star-forming regions or AGNs. On the other hand, filtering out standard emission lines also increases the sensitivity to data artifacts.

    Finally, combining the velocity filter with residual suppression (bottom row) results in a diverse set of anomalies, including both emission-line sources and spectra with moderate continuum deviations.

    A direct comparison of these metrics shows that only 4\% (resp. 14.8\%) of the top 100 (resp. 1000) anomalies are common across all different scores, reinforcing our finding that the anomaly scores are sensitive to different behaviors in spectra, a result consistent with other studies that find low overlap between different anomaly detection methods \citep{reis_2021-unsupervised}, even when only considering variations of the MSE and not completely different metrics.

\subsection{Interpretable Anomaly Detection: Matching Anomalies to Astronomical Intuition} \label{results: representative anomalies}

\begin{figure*}[!ht]
        \centering
        \includegraphics[width=0.95\textwidth]{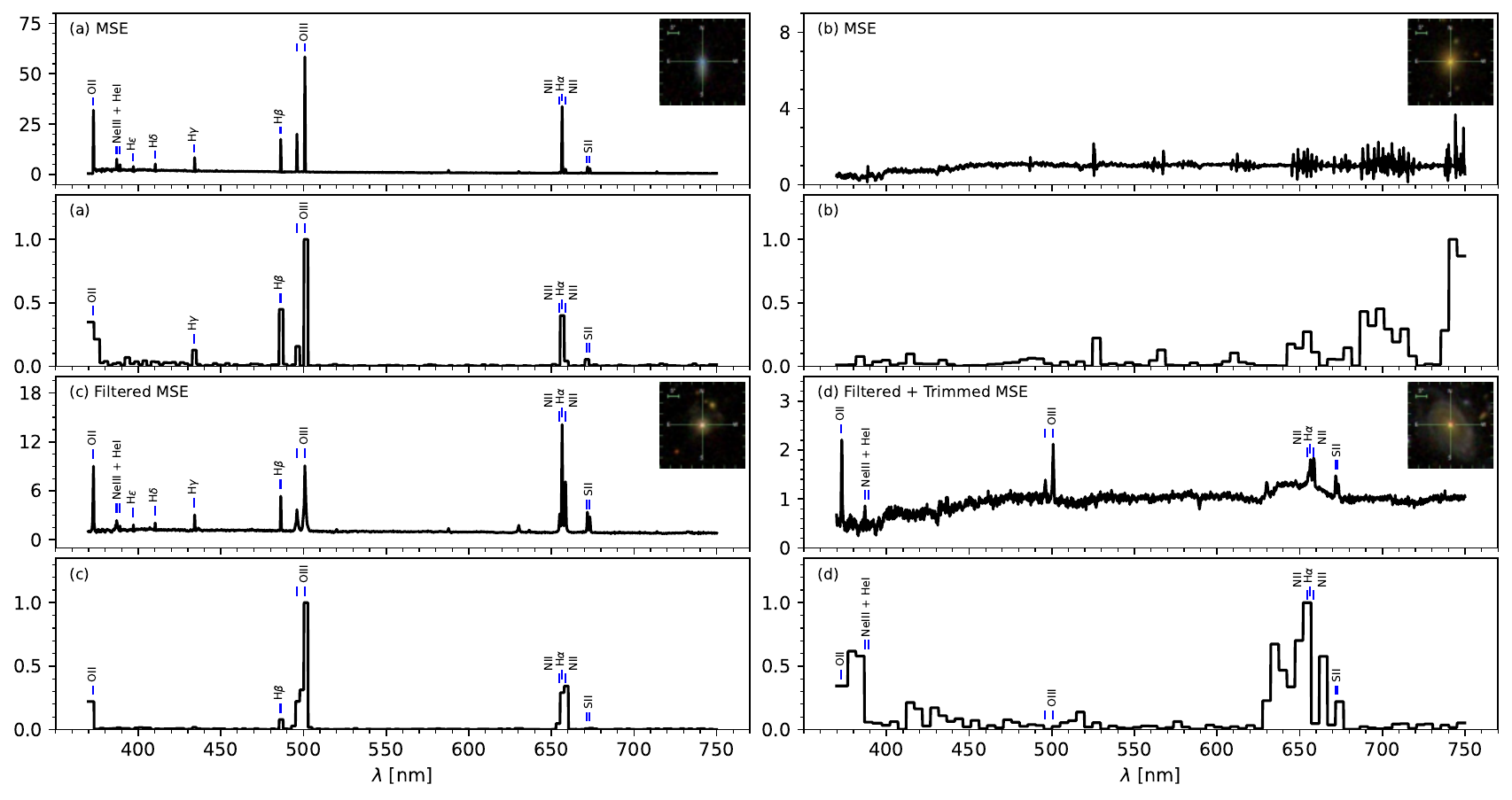}
        \caption{
            Explanations of four representative anomalies detected with different MSE-based scores, each showing the median normalized flux (top, with SDSS imaging thumbnail) and corresponding max normalized LIME explanation weights (bottom).
            Panels illustrate:
            (a) extreme emission-line object driven by [OIII] and H$\alpha$ peaks (MSE);
            (b) red-end weighting from likely instrumental noise (MSE);
            (c) broad emitter with high weights on H$\alpha$ and [OII] (Filtered MSE);
            (d) continuum deviations near H$\alpha$ and the 400~nm break (Filtered + Trimmed MSE).
            Explanations consistently highlight features driving the anomaly, aligning with astronomers' interpretations.
        }
    \label{explanations: mse}
    \end{figure*}

    Beyond identifying outliers, we assess whether the explanation weights provided by our framework align with how an astronomer would interpret the anomalous spectra. To this end, we analyze four of the anomalies illustrated in Fig.~\ref{anomalies: mse-overview} using the explanation weights obtained with our LIME-based approach. The selected spectra and their explanations are shown in Fig.~\ref{explanations: mse}.

    The object in panel~(a) exhibits exceptionally strong and narrow emission lines, with [OII]~$\lambda\lambda$372.7, 372.9, [OIII]~$\lambda\lambda$495.9, 500.7, H$\beta$ and H$\alpha$ dominating the spectrum. These lines reach flux levels $\sim20$ times stronger than the continuum, suggesting the presence of a starburst. The LIME explanation weights plotted below the spectrum confirm this interpretation: the highest weights are sharply peaked at the exact locations of the most prominent emission lines, especially [OII], [OIII], H$\beta$, and H$\alpha$. This outcome illustrates that the anomaly score is driven by physically meaningful features that one would naturally identify as unusual. The interpreter therefore provides a transparent justification for the model's anomaly score decision, reinforcing confidence in the framework's alignment with domain expertise.

    By contrast, the spectrum in panel~(b) highlights a different class of anomalies not tied to extreme astrophysical properties, but instead to data quality issues: a forest of irregular high-frequency fluctuations across the redder wavelengths, with a sharp discontinuity near $\lambda \sim$750~nm. These features lack coherent physical structure and do not correspond to known spectral lines, suggesting the presence of noise or poor sky subtraction. In this case, the LIME weights are concentrated precisely in the problematic red part of the spectrum, confirming that the anomaly score is dominated by the noisy segment.
    This would manually be classified as an observational artifact rather than a genuine astrophysical outlier.
    The fact that the interpreter correctly identifies the region of concern, even in a non-physical case, demonstrates the model and explanation's utility not only in highlighting meaningful features but also in flagging unreliable data products

    Then, in panel~(c), we analyze a spectrum flagged as anomalous using the MSE score with a 250~km~s$^{-1}$ filter applied around the locations of standard emission lines (Table \ref{table: narrow-lines-filter}). At first glance, the spectrum appears dominated by classic nebular emission features such as [OII]~$\lambda\lambda$372.7,372.9, [OIII]~$\lambda$500.7, H$\beta$, and H$\alpha$. Yet what makes this object stand out is not simply the strength of the lines but their unusually broad profiles. The explanation weights reinforce this interpretation: they peak precisely at the locations of these broad features, indicating that the anomaly score is driven not only by flux amplitude but by the width and structure of the lines. This behavior matches expectations, as the applied filter suppresses narrow lines and enhances sensitivity to broader features, such as those associated with turbulent star-forming regions or possible AGN activity.
    The agreement between the score, the explanation, and the astrophysical interpretation offers a compelling validation of the interpreter's ability to highlight relevant anomaly-driving features.

    Finally, in panel~(d), we examine an anomalous spectrum identified using the MSE score filtered at 250~km~s$^{-1}$ and further refined by ignoring the top 3\% of reconstruction residuals from the VAE. This spectrum stands out for its atypical structure: to the left of H$\alpha$, the flux distribution shows irregular broadening, while towards the blue end, just short of the 400~nm break, there is a dip-like flux suppression that deviates from the standard galaxy continuum. The LIME explanation weights confirm that these regions dominate the anomaly score, with high weights concentrated in the vicinity of the 650--670~nm region and in the 380--400~nm range. The ability of the interpreter to isolate these regions helps narrow the interpretation space and guides further scrutiny, an outcome highly aligned with the goals of interpretable anomaly detection.

    These case studies illustrate the value of integrating explanation mechanisms with anomaly detection in spectra. In each example, the LIME-Spectra-Interpreter correctly highlights the spectral regions that an astronomer would identify as responsible for the anomaly, whether due to extreme emission features, broadened line profiles, continuum deviations, or potential data artifacts. This alignment between algorithmic reasoning and expert intuition builds trust in the framework, demonstrating that the model is not only effective at flagging unusual spectra but also capable of offering interpretable justifications that facilitate scientific insight and follow-up analysis.

\subsection{General Trends in Anomalies} \label{results: clustering}

    Sections \ref{results: overview} and \ref{results: representative anomalies} demonstrate that different variations of the MSE-based anomaly score and the LIME-explanation weights are sensitive and highlight diverse anomalous patterns in spectra, including strong emission-line galaxies, unusual continua, and data artifacts. To move from individual case studies to population-level insights, we now analyze global trends using clustering over LIME explanation weights. We focus on the top 1\% most anomalous spectra as measured by the standard MSE score, totalling 1818 spectra.

    Since LIME produces individualized explanations per spectrum, the scale and sign of the weights can vary significantly across the sample. To make these explanations comparable, we focus on the strength of the contribution of each feature using its absolute value and normalize each explanation vector to unit length. This transformation ensures we focus on explanation patterns without being influenced by differences in scale. We use the KMeans clustering algorithm from the scikit-learn\footnote{\url{https://scikit-learn.org/stable/modules/generated/sklearn.cluster.KMeans.html}} library. KMeans is an unsupervised clustering algorithm that partitions data into clusters by minimizing the intra-cluster sum of square Euclidean distances (inertia). It iteratively assigns each point to the nearest cluster centroid and updates centroids based on the mean of assigned points until convergence.

    To obtain an optimal number of clusters, we apply the elbow method, which plots the inertia against the number of clusters and identifies a point from which adding more clusters yields diminishing returns. We also complement the elbow method with the silhouette score, which measures how similar an object is to its own cluster compared to other clusters. A higher silhouette score indicates better-defined clusters. It is worth mentioning that under the unit length normalization of the explanation weights, the Euclidean distance behaves similarly to cosine similarity. Therefore, we minimize the ``curse of dimensionality'' when clustering the explanation weights. We find that seven clusters is adequate for meaningful partitioning of the top 1\% anomalies. We summarize the cluster distribution in Table~\ref{table: cluster-counts-top1}.
    \begin{table}[!ht]
        \centering
        \caption{Distribution of the top 1\% most anomalous spectra across explanation-based clusters.}
        \label{table: cluster-counts-top1}
        \begin{tabular}{crr}
            \hline\hline
            Cluster & Count & Fraction \\
            \hline
            0 & 245 & 13\% \\
            1 & 617 & 34\% \\
            2 & 287 & 16\% \\
            3 & 340 & 19\% \\
            4 & 55  & 3\%  \\
            5 & 196 & 11\% \\
            6 & 78  & 4\%  \\
            \hline
            Total & 1818 & 100\% \\
            \hline
        \end{tabular}
    \end{table}

\begin{figure*}[!ht]
        \centering
        \includegraphics[width=0.95\textwidth]{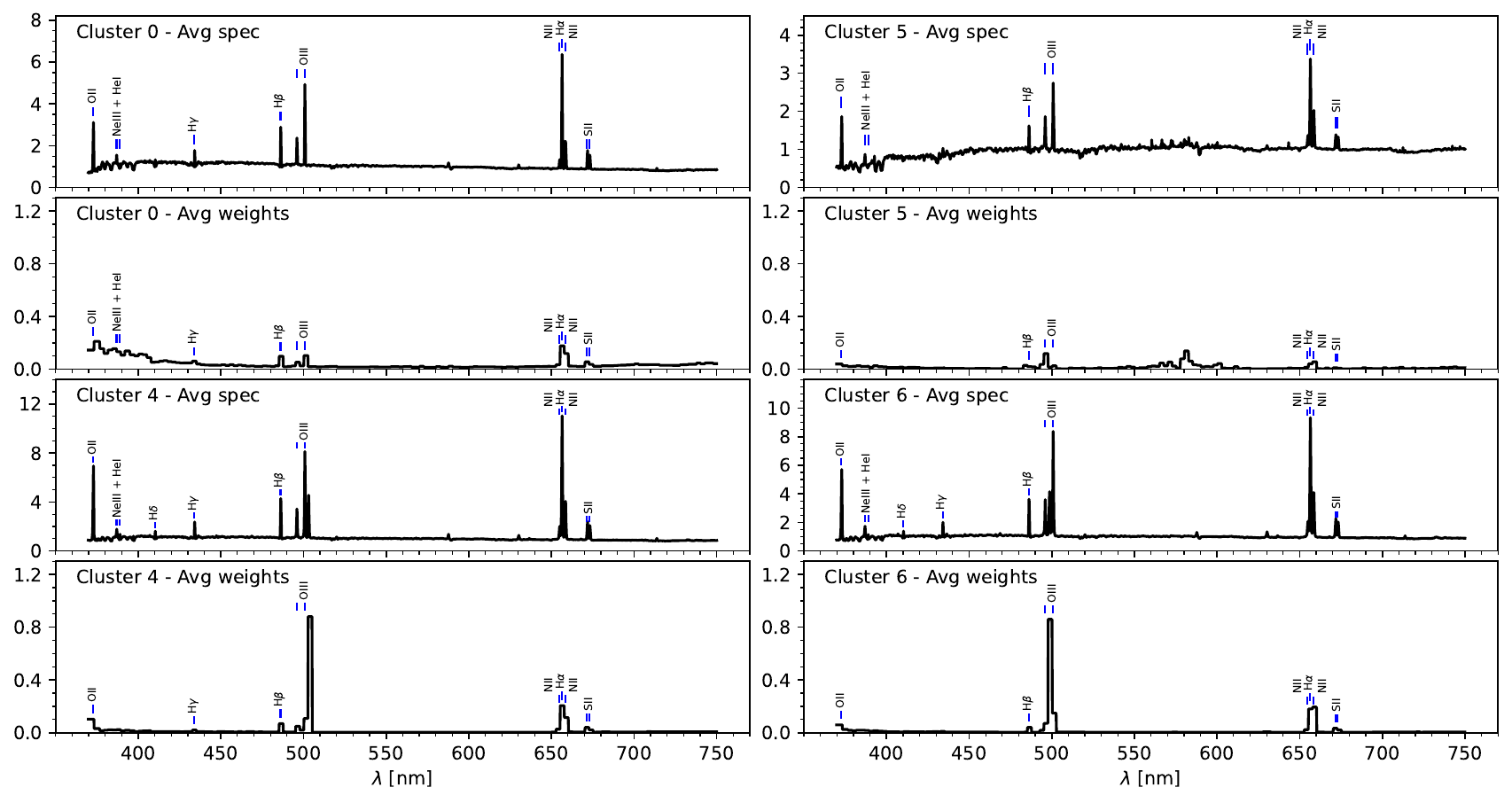}
        \caption{
            Average spectrum and LIME explanation weights for Clusters 0, 5, 4, and 6 from the top 1\% most anomalous spectra (MSE score).
            Clusters 0 and 5 (top rows) show diffuse or noisy explanations with low weights, consistent with poor continuum reconstructions or spikes.
            Clusters 4 and 6 (bottom rows) feature strong weights at truncated [OIII]~$\lambda$500.7 lines due to masked regions during preprocessing.
            This illustrates how explanation-based clustering isolates artifacts from astrophysical signals. Fluxes are median normalized.}
    \label{clustering: artifact-0-5-4-6}
    \end{figure*}

\begin{figure*}[!ht]
        \centering
        \includegraphics[width=0.98\textwidth]{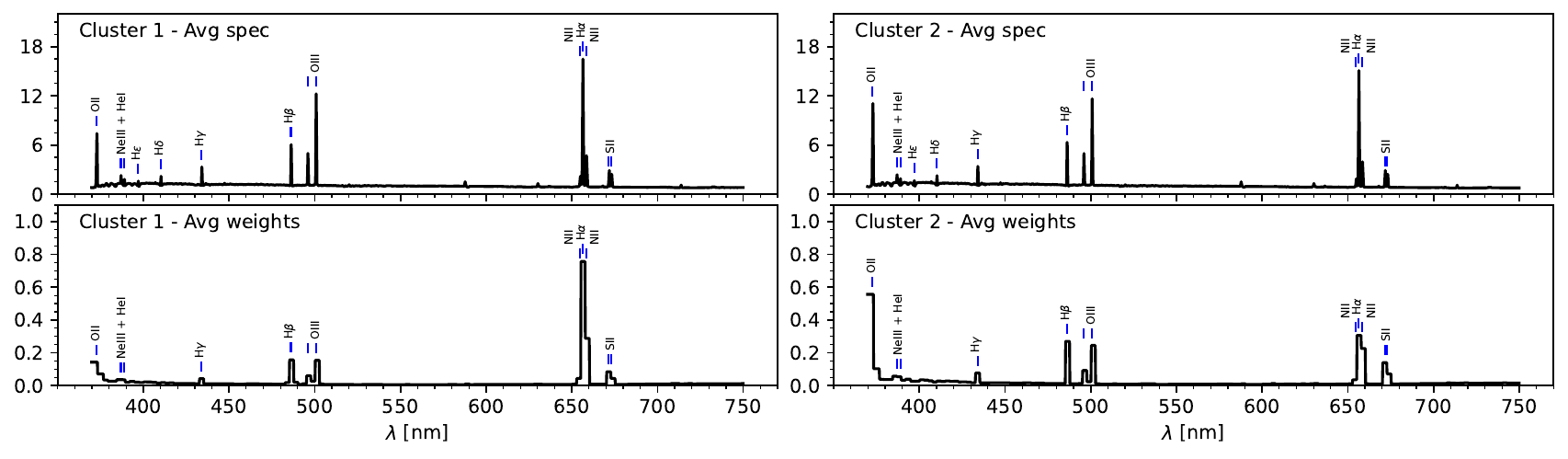}
        \caption{
            Average spectrum and LIME explanation weights for Clusters 1 and 2 of the top 1\% most anomalous spectra (MSE score).
            Cluster 1 (left pannels) emphasizes H$\alpha$+[NII] and [OIII], consistent with dusty, metal-rich starbursts.
            Cluster 2 (right pannels) highlights [OII]$\lambda$372.7, indicating moderate-excitation, enriched H\,II regions.
            Clustering by explanation profiles reveals distinct physical regimes within spectrally similar galaxies. Fluxes are median normalized.}
    \label{clustering: physical-1-2}
    \end{figure*}

\begin{figure}[!ht]
        \centering
        \includegraphics[width=0.98\columnwidth]{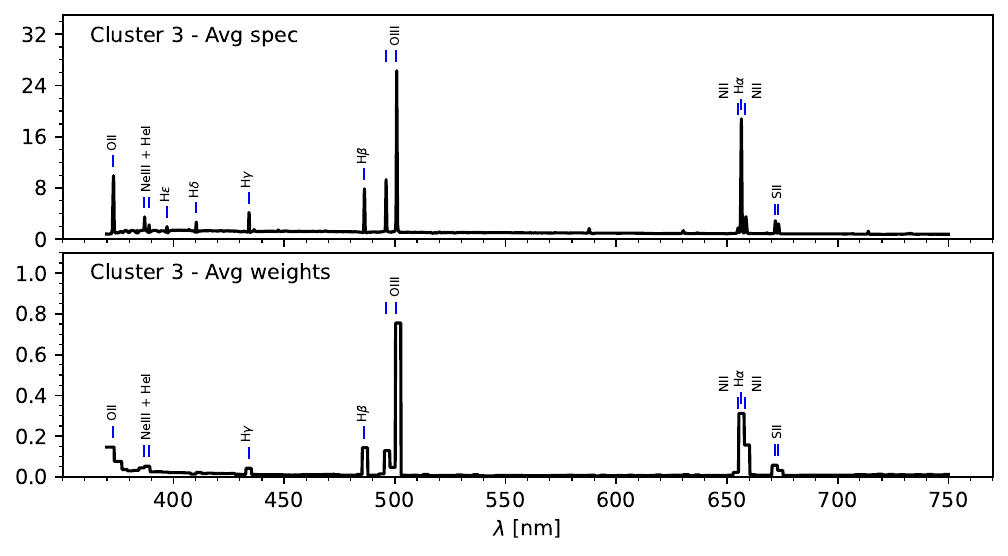}
        \caption{
            Average spectra and LIME explanation weights for Cluster 3 of the top 1\% most anomalous spectra (MSE score).
            Cluster 3 is dominated by [OIII]$\lambda$500.7, characteristic of extreme, low-metallicity systems.
            As discused for clusters 1 and 2, clustering by explanation profiles reveals distinct physical regimes within spectrally similar galaxies. Fluxes are median normalized.
        }
    \label{clustering: physical-3}
    \end{figure}

    We organize our discussion by grouping the clusters into three categories: (i) artifact-driven outliers (clusters 0 and 5), (ii) hybrid cases blending physical signals with data processing artifacts (clusters 4 and 6), and (iii) physically rich emission-line populations (clusters 1, 2, and 3).
    Figures~\ref{clustering: artifact-0-5-4-6}, \ref{clustering: physical-1-2} and \ref{clustering: physical-3} show the average spectrum and explanation for each cluster.

    The inspection of the clustering output reveals that clusters 0 and 5 primarily capture outliers dominated by observational or data reduction artifacts.
    For cluster~0, the left upper panels of Fig.~\ref{clustering: artifact-0-5-4-6} shows that on average these spectra exhibit common emission features such as H$\alpha$, H$\beta$, and [OIII].
    Nonetheless the LIME explanation weights reveal that the anomaly score is not driven solely by these lines.
    Instead, the average explanation shows a broad, uneven structure across the continuum, with an increasing trend in the blue end of the spectrum.
    In fact, the cumulative importance of the region blueward of 450~nm exceeds that of the main emission lines.
    A visual inspection of the top 20 ranked anomalous spectra confirms recurring data artifacts and high-frequency noise around the 400~nm break and the [OII]~$\lambda$372.7 regions.
    A similar pattern emerges in cluster~5.
    We observe in the right upper panels of Fig.~\ref{clustering: artifact-0-5-4-6} that the average explanation is diffuse and fails to target any specific spectral feature.
    A visual inspection of the top-ranked anomalies shows spectra dominated by single sharp spikes at a random wavelengths, with otherwise flat or featureless continua.
    Overall, the outliers in these two clusters are not driven by physical processes in galaxies but rather by data reduction errors or isolated instrumental anomalies.
    This result highlights the discriminative power of the LIME-based framework to isolate systematic outliers, offering an automatic quality control layer for future spectroscopic explorations.

    Clusters~4 and 6, in the left and right bottom panels of Fig.~\ref{clustering: artifact-0-5-4-6}, represent a second category where the detected anomalies reflect a combination of physical structure and data processing artifacts.
    Both clusters are characterized by strong [OIII]~$\lambda$500.7 emission, but this line appears partially clipped across the spectra.
    Upon closer inspection, we identify the origin of this effect: during preprocessing, prior to de-redshifting, the region between 556.5--559.0~nm was masked in all spectra to remove contamination from the strong night-sky line [OI]~$\lambda$557.7.
    Because Clusters~4 and 6 have median redshifts of $z=0.1083$ and $z=0.1181$ respectively,
    the [OIII]~$\lambda$500.7 feature overlaps with this masked region, resulting in systematic truncation.
    The LIME explanation weights consistently assign the highest importance to this region,
    indicating that the clustering successfully identifies a coherent anomaly pattern rooted in preprocessing artifacts.
    While the clipping prevents reliable estimation of excitation line ratios, the spectra in both clusters nonetheless show strong nebular emission and structured continua.
    Photometric inspection reveals predominantly compact morphologies, consistent with small, actively star-forming galaxies.
    This case illustrates how explanation-based clustering can disentangle subtle combinations of physical signals and systematic effects, offering a useful diagnostic for both astrophysical interpretation and data quality control.

    Finally, we now turn to Clusters~1, 2, and 3, which account for 34\%, 16\%, and 19\% of the top 1\% most anomalous spectra, respectively.
    Together, they comprise 69\% of the sample and correspond to physically interesting systems.
    As shown in Fig.~\ref{clustering: physical-1-2} and \ref{clustering: physical-3}, the average spectra in these clusters all exhibit rich nebular emission features, including Balmer series, [OIII]~$\lambda\lambda$495.9, 500.7, [OII]~$\lambda$372.7, and low-ionization lines such as [NII] and [SII].
    While these signatures broadly indicate active star formation, their spectral subtleties and LIME explanation profiles reveal distinct physical regimes across clusters.
    Overall, the structured feature attributions in the LIME explanations point to different line ratios as the key drivers of the anomaly scores.
    Across Clusters~1--3, the Balmer decrement ranges from 3.43 to 3.93 (Table~\ref{table: line diagnostics}), indicating non-negligible dust attenuation.
    \begin{table*}[!ht]
        \centering
        \caption{Emission line diagnostics for Clusters~1--3.}
        \label{table: line diagnostics}
        \begin{tabular}{cccccc}
            \hline
            \hline
            Cluster & 
                \shortstack{Balmer Decrement} & 
                \shortstack{[N\,II]/ H$\alpha$} &
                \shortstack{[O\,III]/ H$\beta$} &
                \shortstack{$[$O\,III$]/$ $[$O\,II$]$} &
                O3N2 \\
            \hline
            1 & 3.74--3.93 & 0.26--0.32 & 1.09--1.40 & 1.06--1.32 & 0.53--0.69\\
            2 & 3.48--3.64 & 0.17--0.21 & 1.58--1.86 & 1.24--1.48 & 0.92--1.02\\
            3 & 3.43--3.54 & 0.11--0.15 & 3.13--3.53 & 2.54--3.06 & 1.23--1.37\\
            \hline
        \end{tabular}
        \tablefoot{
            Table shows key line ratios used in standard classification schemes.
            Each value represents the range between the 45$^{th}$ and 55$^{th}$ percentiles
                across cluster members, based on SDSS pipeline measurements.
            These narrow intervals summarize the central trend while accounting for internal
                variability.
        }
    \end{table*}  

    Cluster~1 (left panels in Fig.~\ref{clustering: physical-1-2}) shows a strong LIME emphasis at H$\alpha$+[NII], with additional attributions (similar) to H$\beta$, [OIII], [OII], and [SII].
    Its line ratios, [NII]/H$\alpha$ (0.26--0.32) and O3N2 (0.53--0.69), suggest moderate metallicity.
    The [OIII]/[OII] ratio (1.06--1.32) indicates a moderately hard radiation field.
    A visual inspection of the SDSS imaging reveals that the top 20 ranked anomalies in this cluster are compact, blue star-forming galaxies.
    This is consistent with the spectroscopic interpretation of this cluster as a population of moderate-excitation, enriched H\,II regions with active star formation.

    Cluster~2 (right panels in Fig.~\ref{clustering: physical-1-2}) exhibits a more line-balanced explanatory pattern dominated by [OII]~$\lambda$372.7, with more evenly distributed weights across other features. The diagnostic ratios, [NII]/H$\alpha$ (0.17--0.21), [OIII]/H$\beta$ (1.58--1.86), and [OIII]/[OII] (1.24--1.48), imply a harder ionizing field and lower metallicities than Cluster~1.
    The SDSS photometry for the top 20 ranked anomalies in this cluster shows a more morphologically-diverse population compared to Cluster 1.
    Several galaxies appear inclined or edge-on, suggestive of disk-like morphologies.
    A number of systems are compact and relatively symmetric, though generally less blue and concentrated than those in Cluster 1.
    Colors vary, with some objects displaying bluer tones indicative of ongoing star formation, while others show redder hues that may reflect older stellar populations.
    In summary, the photometric diversity in Cluster 2 aligns well with its moderate-to-high excitation spectral features. This cluster seems to contain a mix of low-metallicity star-forming spirals and compact galaxies, possibly at more evolved or varied evolutionary stages than those in Cluster 1.

    In contrast, Cluster~3 (Fig.~\ref{clustering: physical-3}) isolates galaxies with stronger excitation conditions compared to previous clusters.
    The average spectrum is dominated by [OIII]~$\lambda$500.7 in both flux and explanation weight.
    Secondary explanation importance is assigned to H$\alpha$+[NII], H$\beta$, [OIII]~$\lambda$495.9, and [OII].
    Its [OIII]/H$\beta$ (3.13--3.53), [OIII]/[OII] (2.54--3.06), and O3N2 (1.23--1.37) ratios exceed those of the other clusters, while [NII]/H$\alpha$ drops (0.11--0.15).
    These values point to low metallicity sytems with hard ionizing fields.
    The SDSS images of the top 20 ranked anomalies in Cluster 3 show predominantly compact, blue-dominated morphologies, consistent with high surface brightness star-forming regions.
    Many exhibit round or slightly irregular shapes, with little evidence of extended structure or disk components.
    A few galaxies display asymmetric light distributions or faint tidal features, potentially indicating interactions or bursts triggered by mergers.
    The uniformity in their compactness and color further supports the interpretation of these systems as low-metallicity, extreme emission-line galaxies, possibly akin to Green Pea analogs or high-ionization starburst galaxies \citep[e.g.][]{cardamone2009}.

    Together, these three clusters demonstrate the power of explanation-driven anomaly detection to uncover astrophysical diversity. Despite their broadly similar emission-line spectra, the LIME-based clustering enables the separation of galaxies into physically coherent subsets, ranging from dusty, moderately enriched bursts (Cluster~1), to lower-metallicity, moderate-excitation systems (Cluster~2), to extreme, low-metallicity emitters with hard ionizing fields (Cluster~3).
    
    To provide a more concrete connection to these galaxies, we present the spectra and corresponding optical imaging for a selection of interesting anomalies from these three clusters in Appendix \ref{appendix: interesting-anomalies}.
    While this work focuses on establishing the methodology and validating it against known astrophysical phenomena, these examples showcase the framework's potential for scientific discovery by highlighting a diverse range of compelling and extreme objects.

\section{Discussion} \label{section: discussion}
\subsection{Interpretation speed}

    The latest generation of large surveys in astronomy will generate massive amounts of data and produce numerous alerts as they find anomalies in near real time.
    Tools like the one presented in this work will be increasingly essential to explore such alerts efficiently, helping astronomers triage and interpret them. To illustrate the practical feasibility of our method, we measured the explanation generation time: on a standard notebook with an AMD Ryzen 7 PRO 5850U CPU, LIME produces one explanation in approximately 0.82~s per core. Scaling this to a dataset of 1 million spectra results in a total runtime of roughly 9.5 days on a single core. However, with basic parallelization, e.g., 100 parallel jobs on a small computing cluster, this time drops to $\sim$2.3 hours. These results show that our interpretability framework is scalable and well-suited for integration into high-throughput pipelines.

\subsection{Effects of segmentation and perturbation}\label{section: explanation-config}

    LIME explanations depend on several hyperparameters, including the type of perturbation, the segmentation strategy, the number of segments, and the number of perturbed samples used to fit the local surrogate model. These choices can significantly affect the resulting explanations. In the remainder of this section, we systematically assess the sensitivity of LIME explanations to each parameter by varying them independently while keeping the others fixed to their default values (Sect.~\ref{section: iml}).

\subsubsection{Perturbation}

\begin{figure}[!ht]
        \centering
        \includegraphics[width=0.98\columnwidth]{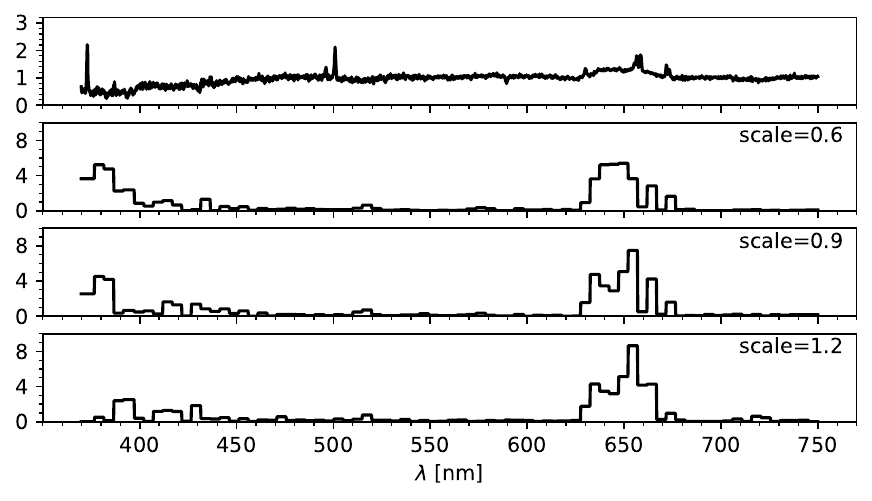}
        \caption{
            Explanation weights under different flux scaling perturbations of 0.6, 0.9 (default), and 1.2.
            The overall structure is stable across scale factors.
            Fluxes are median normalized and weights are factors of 10$^{-2}$.
        }
        \label{params: scale-variation}
\end{figure}

    The perturbation strategy plays a critical role in how LIME assigns explanation weights to different spectral regions. In this work, our default approach is to apply a flux scaling within randomly selected segments. This perturbation preserves local flux structures, aligning with the interpolation capabilities of neural networks. We find this method to be robust across a range of scaling factors. Fig.~\ref{params: scale-variation} shows the explanation weights for a spectrum when applying scaling factors of 0.6, 0.9 (default), and 1.2. The overall attribution pattern remains qualitatively consistent, particularly in the continuum. This suggests that the explanations are stable against moderate changes in scaling magnitude.

\begin{figure}[!ht]
        \centering
        \includegraphics[width=0.98\columnwidth]{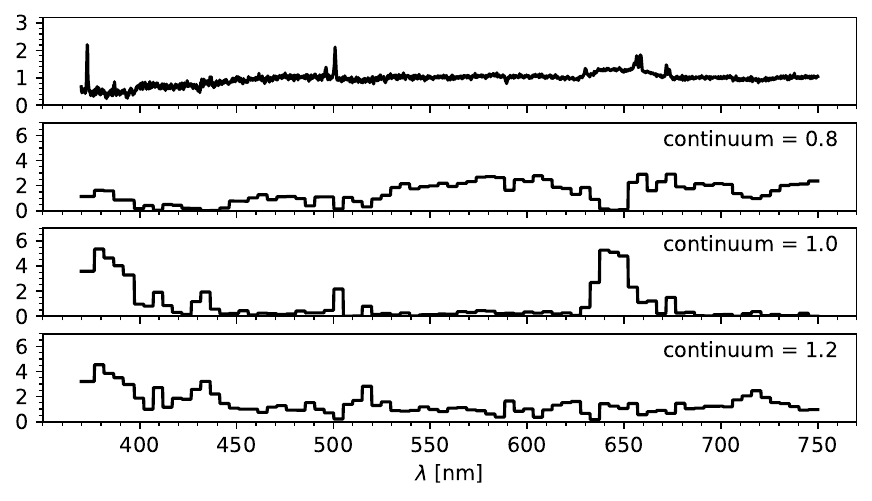}
        \caption{Explanation weights using flat perturbation values of 0.8, 1.0, and 1.2. Using 1.0 (third panel)--the median-normalized flux--yields coherent results, while deviations (second and fourth panels) introduce spurious continuum weights.
        Fluxes are median normalized and weights are factors of 10$^{-2}$.}
        \label{params: flat-variation}
    \end{figure}

    We then test flat perturbations, where each segment is replaced by a constant value to ``gray out'' its signal. Given that all the spectra are median-normalized, a natural baseline is 1.0. Figure~\ref{params: flat-variation} compares this to values of 0.8 and 1.2, showing that small deviations introduce spurious continuum attributions. In contrast, using 1.0 yields results closest to the scaling-based explanations, confirming it as the most neutral choice.

\begin{figure*}[!ht]
        \centering
        \includegraphics[width=0.98\textwidth]{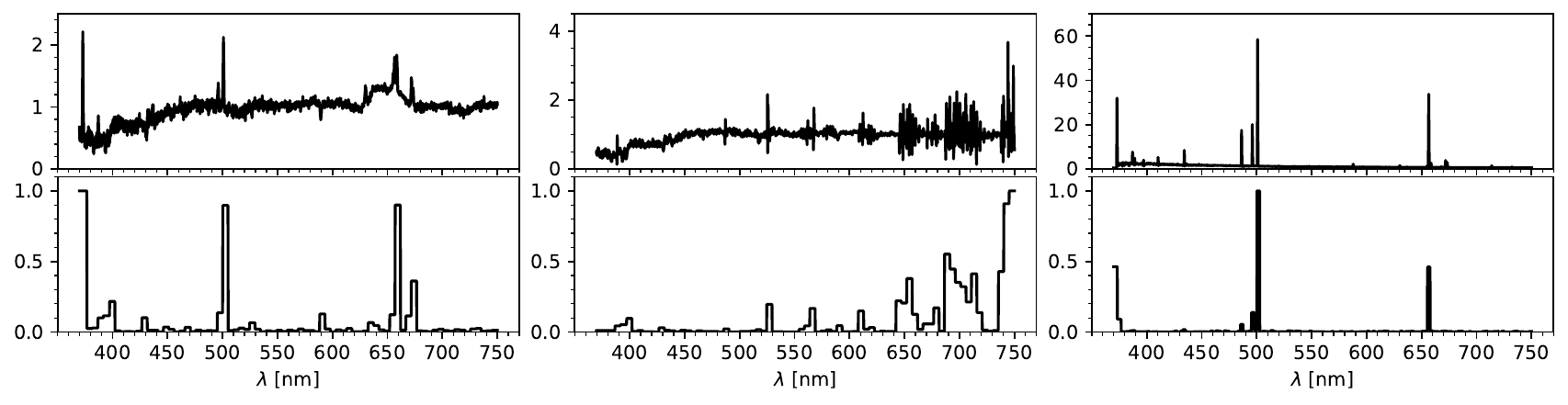}
        \caption{Explanation weights using a segment's mean flux as perturbation. Left: example spectrum from earlier sections. Middle: artifact-dominated object. Right: strong-emission-line galaxy. While this method highlights prominent lines effectively, it may also amplify flux gaps or downweight continua. Segmentation uses 77 segments (left, middle) and 150 segments (right).
        Fluxes are median normalized and weights are max normalize.}
        \label{params: with-mean-variation}
    \end{figure*}

    We also test a variant where each segment is replaced by its mean flux, removing local structure while preserving the global level. The left panel of Fig.~\ref{params: with-mean-variation} shows that this perturbation highlights sharp emission and absorption lines, while continuum regions receive fewer attributions. There is an explanation peak around 517.5~nm arising from missing data, illustrating this method's sensitivity to flux contrasts. To further explore this behavior, we apply this scheme to a spectrum dominated by sky-subtraction artifacts and another with strong emission lines (middle and right panel in Fig.~\ref{params: with-mean-variation}). In both cases, the explanations correctly highlight the unusual features in each case, but broader attribution patterns are more difficult to interpret. This method can be helpful in diagnosing artifact-dominated spectra or probing flux ratios, but it requires caution.

    In summary, the perturbation method has a substantial impact on the resulting explanations. Scaling-based perturbations provide stable, interpretable results that preserve local structure and are most consistent with the model's training distribution. Flat or mean-based perturbations can be informative in specific contexts but risk injecting artifacts or suppressing physically meaningful features.

\subsubsection{Segmentation}

\begin{figure*}[!ht]
        \centering
        \includegraphics[width=0.95\textwidth]{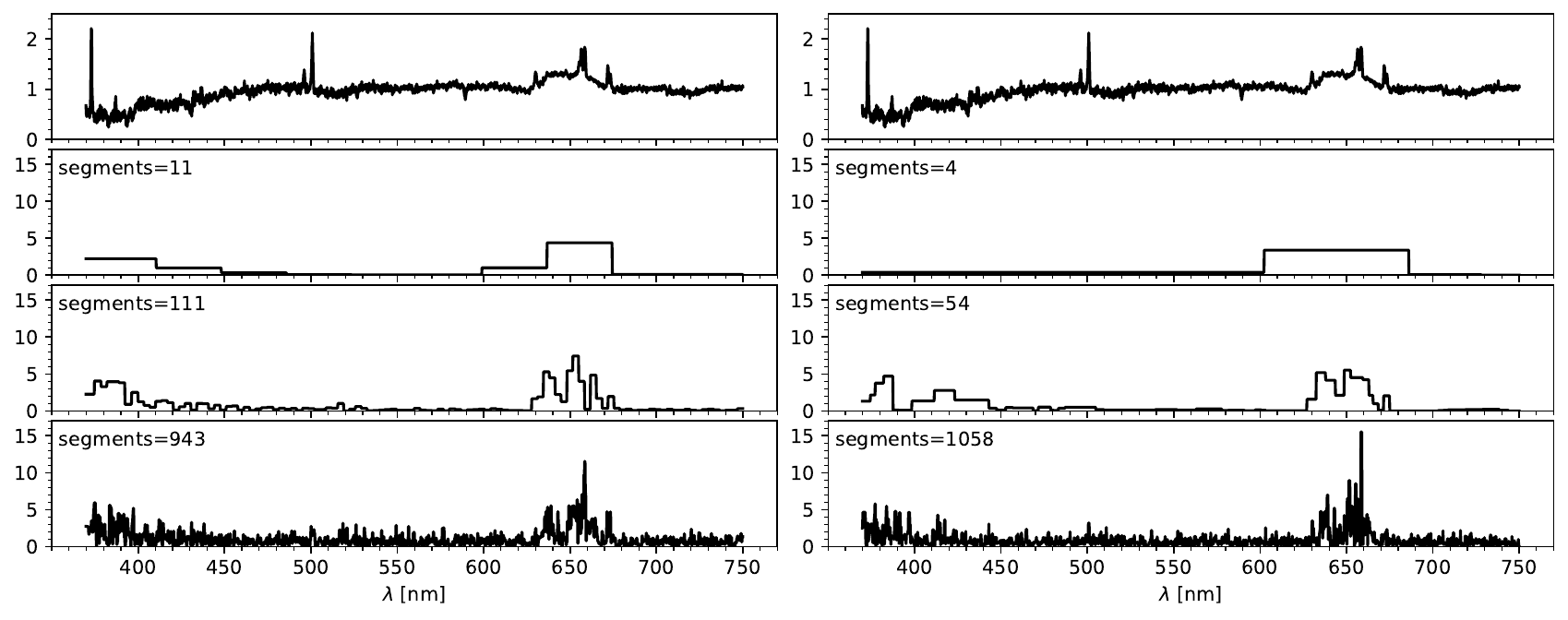}
        \caption{
            Effect of segment count on explanation quality.
            Left: uniform segmentation with 11, 111, and 943 segments.
            Right: SLIC segmentation with equivalent upper bounds, yielding 4, 54, and 1058 segments.
            Having too few segments dilutes key features; having too many adds noise.
            Intermediate settings yield clearer and more meaningful explanations.
        Fluxes are median normalized and weights are factors of 10$^{-2}$.}
        \label{params: number-segments}
    \end{figure*}

    Two key parameters govern how the input spectrum is segmented for perturbation-based explanation: the number of segments and the number of perturbed samples. These parameters are interdependent, as the number of segments defines the dimensionality of the local neighborhood that the interpreter must sample. Specifically, given $n$ segments, the number of possible perturbed combinations is $2^n - 1$. A small number of segments results in fewer degrees of freedom and requires fewer samples to fit the surrogate model. Conversely, a larger number of segments expands the combinatorial space, demanding more samples to faithfully approximate the model's local behavior. However, fewer segments often group together physically distinct regions (e.g., continuum and emission lines), potentially washing out interpretability. At the other extreme, too many segments may lead to overfitting or include segment sizes smaller than the spectral resolution, an ill-posed scenario. Figure~\ref{params: number-segments} illustrates these trade-offs using both uniform segmentation (left panel) and SLIC segmentation (right panel). The uniform segmentation divides the spectrum into equal-width segments, while SLIC uses a clustering approach to adaptively determine segment boundaries based on flux structure.

    In both segmentation schemes, explanations with very few segments tend to concentrate weights over broad continuum regions, failing to isolate specific features. With many segments (e.g., > 942), explanations become noisy and less structured. Notably, the SLIC segmentation with an upper bound of 111--yielding 54 segments--produces an explanation that captures the anomalous continuum structure of this spectrum. It is interesting to note that the SLIC segmentation with and upper bound of 943 segments yields 1058 segments, which is more than the number of segments in the uniform segmentation. This results in a very noisy explanation, as the segments are too small to capture meaningful flux variations.

\begin{figure}[!ht]
        \centering
        \includegraphics[width=0.98\columnwidth]{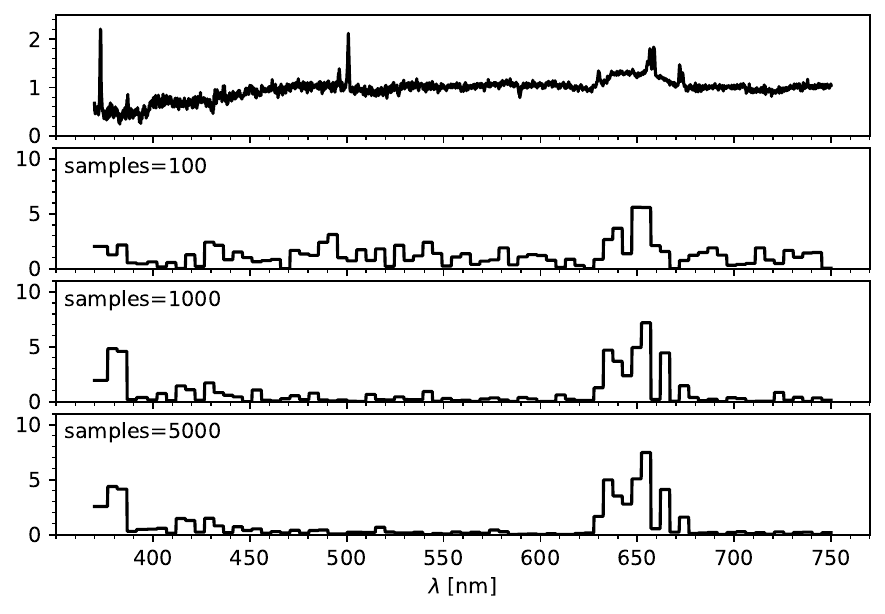}
        \caption{
            Effects of the number of perturbed samples on the quality of the explanation.
            Explanations stabilize and improve as more perturbed samples are used.
            Too few samples yield noisy or misleading weights.
            Fluxes are median normalized and weights are factors of 10$^{-2}$.
        }
        \label{params: number-samples}
    \end{figure}

    We also explore how the number of perturbed samples affects explanation quality. Fig.~\ref{params: number-samples} shows explanations using 100, 1000, and 5000 perturbed spectra under a uniform segmentation of 77 segments. With only 100 samples, the interpreter produces noisy and unreliable weights. At 1000 samples, the structure begins to emerge with a small continuum noise. Finally, at 5000 samples the explanation is stable and aligns with the anomalous nature of this spectrum. This emphasizes that under high-dimensional perturbation spaces, sufficient sampling is essential to suppress artifacts and recover faithful attributions.

\subsection{Interpretable machine learning in astronomy}

    As machine learning becomes ubiquitous in astronomy, extending interpretability frameworks to accommodate different astronomical data types is increasingly important. This work demonstrates how model-agnostic explainability tools such as LIME can be successfully adapted to spectroscopic data to provide physically grounded insights into anomaly detection results. In our case, two core components were adapted: segmentation and perturbation. Instead of pixel-based superpixels used in image tasks, we segment galaxy spectra into wavelength intervals, uniformly or via clustering. Perturbations, rather than masking or replacing features, involve flux rescaling within segments, preserving spectral structure in a way compatible with anomaly detection. This modularity highlights LIME's broader flexibility. The same framework could, in principle, be extended to other astronomical domains. For instance, time series can be sliced into temporal segments, akin to spectral segments. Photometric data can be explained using passbands or color indices as interpretable features. Data cubes, such as IFU (Integral Field Unit) observations, present a more complex case but can be segmented spatially and spectrally to assess local and global contributions to predictions. These generalizations are especially relevant in the context of ongoing and upcoming large-scale surveys such as LSST, DESI, and 4MOST, which, combined, will produce massive volumes of multi-modal data requiring scalable and interpretable analysis tools.

    In this paper, we have shown that interpretable anomaly detection is not just viable but powerful: explanation weights align with astrophysical intuition, and clustering of explanations reveals structured trends in the top anomalies.
    This opens the door for automated prioritization, classification, and debugging workflows that are transparent and scientifically meaningful.

\section{Conclusions}\label{section: conclusions}

    In this work, we introduced a framework for interpretable anomaly detection of SDSS galaxy spectra using a variational autoencoder (VAE) coupled with a customized LIME interpreter. Our key contributions include: (1) A flexible VAE-based anomaly scoring system using multiple reconstruction-based metrics, each capturing distinct types of spectral deviations. (2) A spectral adaptation of LIME that segments spectra into interpretable wavelength regions and applies perturbations through flux scaling, enabling local explanations of anomaly scores. (3) A systematic analysis of segmentation and perturbation choices, with default parameters selected to balance resolution and stability. (4) Clustering of LIME explanation weights for the top 1\% most anomalous spectra, revealing astrophysically meaningful subgroups aligned with known emission line diagnostics and morphological traits.

    By providing an interpretation of why spectra are anomalous, our framework connects the complex outputs from our model with expert-driven astrophysical insights. 
    The explanations isolate features such as strong emission lines, continuum irregularities, and instrumental artifacts, enabling automatic categorization and quality control. Our clustering analysis further demonstrates that the explanation space encodes rich physical structure, distinguishing dusty starbursts, chemically enriched H\,II regions, and extreme low-metallicity emitters.
    This work illustrates how interpretable ML can scale to large spectroscopic datasets and deliver insights beyond an anomaly score, guiding both follow-up science and model development. The methodology is generalizable to other domains in astronomy, supporting the broader integration of explainability in the data-driven era of astronomy.

    \section*{Code availability}

    All code used in this work is publicly available under an open-source license and can be found at the GitHub link in the footnote\footnote{\url{https://github.com/ed-ortizm/Interpreting-Anomaly-Detection-in-SDSS-Spectra}}. The organization hosts modular repositories for each component: data processing (sdss), VAE training (autoencoders), anomaly scoring (anomaly), and LIME-based explanation of spectra (xai-astronomy). Code is written in Python, with dependencies and usage instructions provided in each repository to ensure reproducibility.

    \begin{acknowledgements} 

    EO belongs to the PhD Program \textit{Doctorado en Física, mención Física-Matemática}, Universidad de Antofagasta, Antofagasta, Chile, and acknowledges support from the National Agency for Research and Development (ANID)/Scholarship Program/DOCTORADO BECAS NACIONAL CHILE/2018-21190387.
    MB acknowledges support from the ANID BASAL project FB210003. This work was supported by the French government through the France 2030 investment plan managed by the National Research Agency (ANR), as part of the Initiative of Excellence of Université Côte d’Azur under reference number ANR-15-IDEX-01.

    \end{acknowledgements} 
    
    \bibliographystyle{aa}
    \bibliography{bibliography}
    \begin{appendix}

        \section{VAE architecture} \label{appendix: VAE}

        The architecture of a VAE is defined with the number of layers in the encoder, the decoder, and the dimensionality of latent space. Each layer is defined by a set of neurons. The input of a neuron is the output of the neurons in the previous layer. The neuron performs a dot product of its inputs with a vector of weights, then it adds a bias term. The results of these operations is the input of an activation function. In this work we use the standard Rectified Linear Unit (ReLU) function,
        \begin{eqnarray}
            \text{ReLU}(x) = \text{max}\{0, x\},
        \end{eqnarray}
        where max represents the maximum value between 0 and $x$. The output of a neuron is the output of the activation function.

        We trained our VAE following standard procedures \citep{geron2017, patel2019}. Shortly, we define a loss function, a training and a validation set, and a criterion to stop training. The loss function guides the learning process of the VAE during the training phase. During training, batches of spectra from the training set are feed to the VAE. The reconstructions and latent representations of each batch are used to compute the loss function. If the loss function is not reduced, then the parameters of the VAE are updated using the back propagation algorithm \citep{geron2017, patel2019}. In this work, we used the Adam optimization algorithm for back propagation. The algorithm completes a training epoch when all the batches accounting for the training are feed to the model. The training finishes when the value of the loss function on the validation set does not drop during 10 epochs. It is worth remarking that the validation set contains different spectra from the training set, and it is used as a control set to avoid overfitting of the training set. In this work we train our VAE with spectra from SDSS DR16, see Sect.~\ref{section: data}. The training and validation set contain 80\% and 20\% of the spectra of a given bin, respectively.

        The loss function is given by the following equation,
        \begin{eqnarray}
            \label{eq: infoVAE}
            \mathcal{L} \big(X, X', w\big)
            &=& \alpha \cdot \text{MSE}\big( X,X' \big)
            + 
            \text{D}_{\text{KL}}\big(q(w|X)\hspace{.5mm}||\hspace{.5mm}p(w)\big)
            \\ \nonumber
            &+& (\lambda - 1)
            \cdot
            \text{MMD} \big(q(w|X)\hspace{.5mm}||\hspace{.5mm} p(w)\big).
        \end{eqnarray}

        The loss function in Eq.~\ref{eq: infoVAE} contains three terms. A reconstruction term defined by the Mean Squared Error,
        \begin{equation}
            \label{eq: mse}
            \text{MSE} \big( X, X' \big) =
            \frac{1}{N} \sum_i^N \left(x_i - x_i'\right)^2,
        \end{equation}
        where $N$ represents the number of fluxes in a spectrum, denoted by $X$,and $x_i$ denotes a single flux of $X$. The MSE drives the VAE reconstruction power and helps to encode similar spectra onto clusters in the latent space \citep{shafkat2018}.

        The second term is the Kullback-Leibler Divergence (KLD) \citep{kingma2013},
        \begin{equation}
            \label{eq: KLD}
            \text{D}_{\text{KL}} \big(q(w)\hspace{.5mm}||\hspace{.5mm}p(w)\big)
            =
            \mathbb{E}_{q(w)}
            \left[ -\log \left(\frac{q(w)}{p(w)} \right) \right],
        \end{equation}
        where $q(w)$ and $p(w)$ are the probability distribution of the encoder and the prior for the decoder respectively. Finally, $\mathbb{E}_{q(w)}$ is the expected value over $q(w)$.

        The last term is Maximum Mean Discrepancy (MMD) \citep{zhaoShegjia2017},
        \begin{eqnarray}
            \label{eq: MMD}
            \text{MMD} \big( q(w) \hspace{.5mm}||\hspace{.5mm} p(w) \big)
            &=&
            \mathbb{E}_{p(w), p(w')}
                \left[k(w, w') \right] \\ \nonumber
                &+& \mathbb{E}_{q(w), q(w')} \left[ k(w, w') \right]
                \\ \nonumber
                &-& 2 \mathbb{E}_{p(w), q(w')} \left[ k(w, w') \right],
        \end{eqnarray}
        where $w$ is the latent representation of a spectrum and $w'$ is a vector with the same dimension of $w$ but sampled from prior of the decoder. 

        The KLD and MDD act as regularization terms to improve smoothness on the latent space \citep{zhaoShegjia2017, shafkat2018}. A smooth latent space is what makes VAEs generative models, since they allow a meaningful interpolation among points in it. A VAE with only the MSE, is an Auto Encoder (AE). An AE does not guarantee that an interpolation over empty regions in latent space will generate realistic spectra, therefore limiting its capabilities. This is because these gaps contain points not seen by the model during the training phase. To alleviate this situation, in conjunction with the MMD and KLD terms, VAEs incorporate a stochastic layer. This layer samples each spectrum over a distribution in the latent space during training. For this work, we use a multivariate Gaussian distribution, where its means and variances are trainable parameters of the model. In the training process, the VAE will see each spectrum several times. Therefore, thanks to this layer, each point will be sampled over a small region of latent space, rather than to a single point. Given this stochastic layer, the encoder network constitutes a distribution on the latent space, that is, the encoding distribution. As mentioned before, this distribution is represented by $q(w)$, and the prior for the decoder by $p(w)$. The objective of the KLD term is to make the encoder distribution in latent space similar to that of a normal distribution in a point by point fashion. Henceforth, the latent representation of the data is encouraged to be evenly distributed around the origin of the latent space. Finally, the MMD term, discourages the KL divergence to force the encoding distribution to match the prior, avoiding an uninformative latent space. The MMD term accomplishes this objective by encouraging the encoding distribution to match the prior $p(w)$ of the decoder in expectation \citep{chenXi2017, zhaoShegjia2017}. This is accomplished using a kernel function $k(w,w')$ as shown in Eq. \ref{eq: MMD}. The kernel function measures the similarity between two samples, in this case, a sample of $w$ from the encoding distribution and sample of $w'$ from the prior of the decoder, in our case a multivariate normal distribution. In this work we use a Gaussian kernel,
        \begin{equation}
            \label{eq: gaussian-kernel}
            k \big( w, w' \big) =
            e^{
                - \frac{|| \hspace{.5mm} w-w'||^2}{2 \sigma^2}
            },
        \end{equation}
        where the variance of the kernel was defined as the inverse of the dimensionality in the latent space, following recommendations in \cite{zhaoShegjia2017}.

        To select our models, first we defined different architectures, and for each we performed a grid search for $\alpha$ and $\lambda$. Among the different architectures, we found that the model with the best reconstructions have three layers in the encoder and the decoder and a latent dimension of 12. The number of neurons per layer, including the latent space, is given by $256-128-64-12-64-128-256$. The values of $\alpha$ and $\lambda$ for the best performing model trained on the bin with the largest SNR are 0 and 8.6 respectively.
        
        \newpage
        \section{Lines}
        \begin{table}[!ht]
            \caption{Narrow emission lines used in filters for anomaly scores}
            \label{table: narrow-lines-filter}
            \centering
            \begin{tabular}{cc}
                \hline\hline
                Line & Wavelength   \\
                {}   & (nm)        \\
                \hline
                OII          & 372.6040 \\
                $H_{\delta}$ & 410.1734 \\
                $H_{\gamma}$ & 434.0472 \\
                $H_{\beta}$  & 486.1352 \\
                OIII         & 495.8911 \\
                OIII         & 500.6843 \\
                NII          & 654.8041 \\
                $H_{\alpha}$ & 656.2787 \\
                NII          & 658.3461 \\
                SII          & 671.6440 \\
                SII          & 673.0812 \\
                \hline
            \end{tabular}
        \end{table}
        \section{Interesting anomalies from Cluster analysis} \label{appendix: interesting-anomalies}
        
        This appendix showcases a selection of representative interesting anomalies from the three main
        physically-driven clusters discussed in section \ref{results: clustering}.
        For each of the three main clusters, we present four representative galaxies to connect clusters'
        properties derived from the LIME weights with their spectral characteristics, the features that drive
        the anomaly score, with the underlying physical state and morphology revealed by their optical imaging.

\begin{figure}[!ht]
            \centering
            \includegraphics[width=0.99\columnwidth]{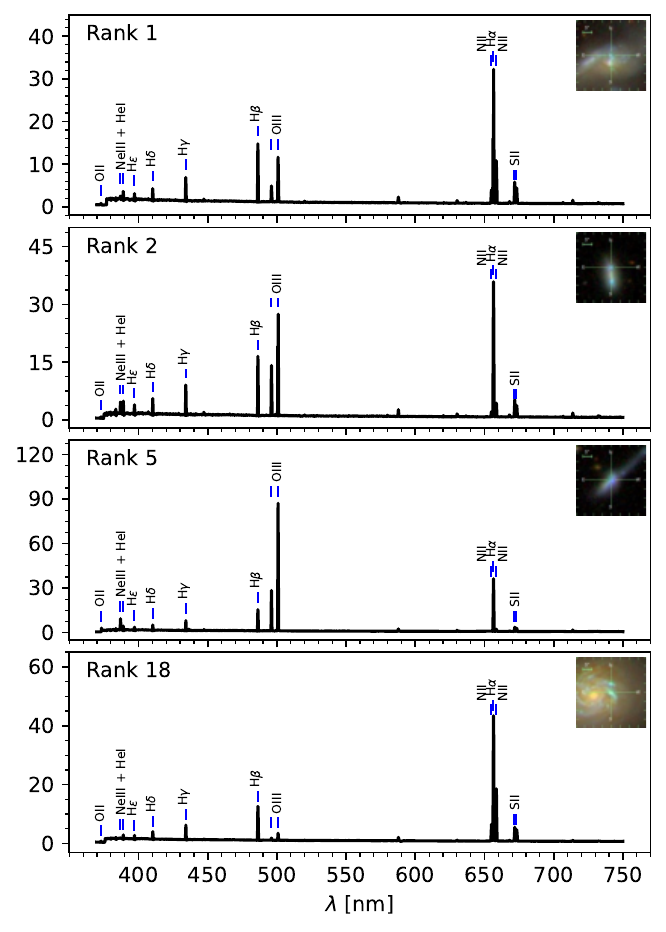}
        \caption{
            A selection of four representative anomalies from Cluster 1,
            which primarily consists of moderate-excitation, enriched H II regions. 
            The sample highlights the diverse morphologies linked to this spectral class, including
            a disturbed spiral (Rank 1), an edge-on starburst (Rank 2), a compact galaxy (Rank 5),
            and an ongoing merger (Rank 18).
            Fluxes are median normalized.
        }
        \label{cluster1: anomalies}
        \end{figure}

        The galaxies presented in Figure~\ref{cluster1: anomalies} are examples from cluster~1,
        confirming its nature as a collection of moderate-excitation, enriched H II regions.
        All four spectra showcase the key features identified from the cluster's average properties:
        strong Balmer emission lines (H$\delta$, H$\gamma$, H$\beta$) and a prominent H$\alpha$+[NII] that is the most
        dominant feature, consistent with the average LIME weights that peak in this region.
        This shared spectral signature points to a common physical driver of intense star formation.
        Interestingly, the examination of these individual anomalies reveals important nuances.
        While Ranks 1, 2, and 18 are archetypal members of this cluster with strong H$\alpha$ emission and moderate
        [OIII]/H$\beta$ ratios, Rank 5 stands out as a more extreme case.
        Its spectrum shows the strongest emission lines in this small sample, with the [OIII] line being significantly stronger than H$\alpha$, indicating a considerably higher ionization parameter than is typical for this group.
        This highlights the diversity within a single cluster and demonstrates how the method groups objects with broadly similar interpretation patterns, even with significant variations in line strength.
        The photometry in this sample provides additional context.
        The thumbnails in the upper right of each spectrum reveal that these class of anomalies are found in a variety
        of dynamically active systems, from the disturbed spiral of Rank 1 and a starburst in the spiral arms of
        Rank 18 to the clumpy edge-on disk of Rank 2.
        The particularly high-excitation spectrum of Rank 5 corresponds to an intensely blue and compact morphology, suggesting a concentrated and powerful starburst event.

\begin{figure*}[!ht]
            \centering
            \includegraphics[width=0.95\textwidth]{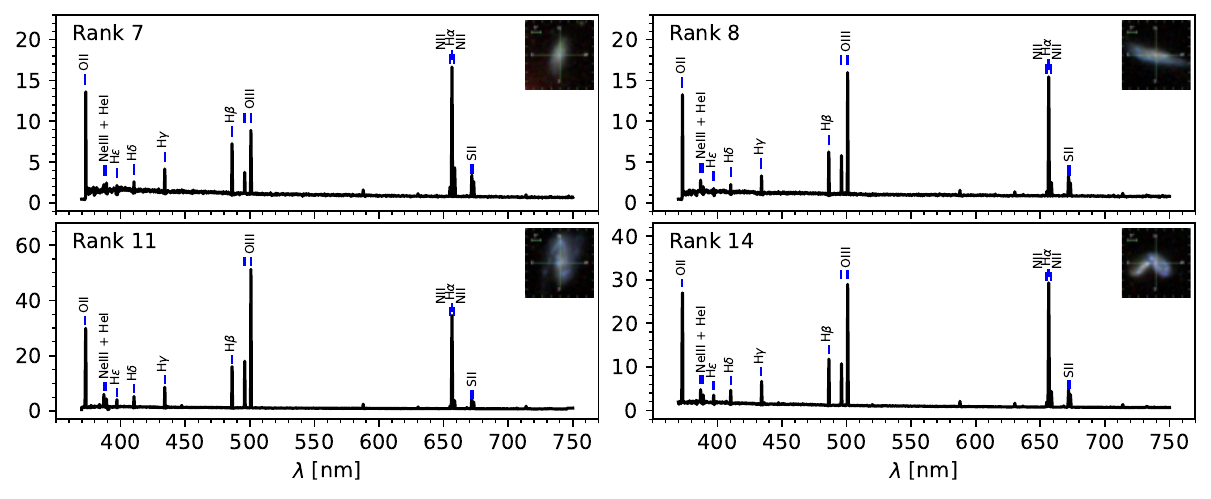}
        \caption{
            A selection of four representative anomalies from Cluster 2,
            characterized by harder ionizing fields and lower metallicities than Cluster 1.
            This cluster is characterized by a wide range of interaction-driven morphologies,
            including a ``tadpole'' galaxy (Rank 7), a warped disk (Rank 8),
            a chaotic merger (Rank 11), and a post-merger with tidal tails (Rank 14).
            Fluxes are median normalized.
        }
        \label{cluster2: anomalies}
        \end{figure*}

        The anomalies in cluster~2 (Fig~\ref{cluster2: anomalies}) represent a distinct population with a
        significantly higher ionization state compared to cluster~1.
        This is spectrally evident in two main ways across the examples.
        First, the [OIII] emission becomes much more prominent, being of similar strength, and in some cases stronger
        than H$\alpha$, as seen in the spectrum of Rank 11.
        This highlights the harder ionizing field characteristic of the cluster.
        Second, another key feature is the consistently strong [OII] emission, which aligns with the average
        LIME interpretation for this group being centered on this line. 
        The photometry of these objects shows morphological evidence of gravitational interactions.
        This is evidenced by a diverse range of features, including the asymmetric ``tadpole'' structure of Rank 7,
        the large-scale warp in the disk of Rank 8, the chaotic morphology in Rank 11 pointing to a merger,
        and the tidal tails of the post-merger system in Rank 14.

\begin{figure*}[!ht]
            \centering
            \includegraphics[width=0.95\textwidth]{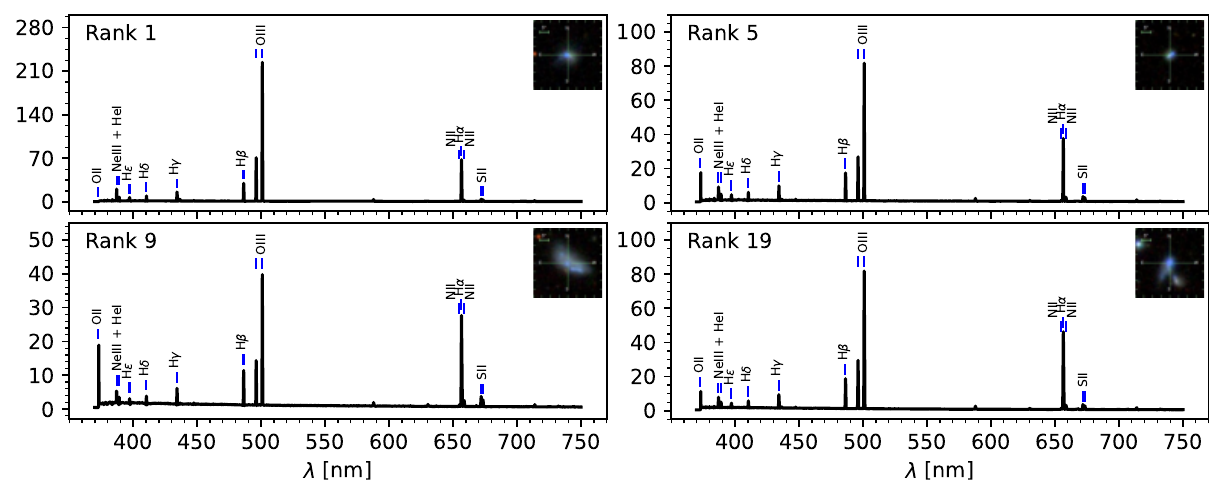}
        \caption{
            A selection of four representative anomalies from cluster~3, which isolates extreme emission-line,
            low-metallicity galaxies resembling ``Green Pea'' analogs.
            The examples showcase the compact, blue morphologies characteristic of this class,
            often found within interacting or merging systems.
            Fluxes are median normalized.
        }
        \label{cluster3: anomalies}
        \end{figure*}

        Finally, cluster~3 (Fig.~\ref{cluster3: anomalies}) represents the most spectrally extreme objects identified
        by our framework, consistent with the properties of rare ``Green Pea'' galaxies.
        All four spectra perfectly embody the defining characteristic of this class:
        they are overwhelmingly dominated by the [OIII] $\lambda\lambda$495.9,500.7~nm emission lines.
        This is a direct confirmation of the average LIME interpretation for this cluster,
        which peaks strongly on this feature.
        Furthermore, the characteristically low [NII]/H$\alpha$ ratio, visible in all four panels,
        is a classic signature of the low-metallicity gas and hard ionizing radiation fields that define these systems.
        Photometry reveals that these extreme spectra originate in intensely blue and compact galaxies,
        with Rank 1 and 5 serving as archetypal examples. Rank 9 shows a chaotic merger while
        Rank 19 shows an interacting close pair of galaxies, suggesting that these extreme events are merger-triggered.


        Taken together, these examples illustrate interesting anomalous spectra and one of the utilities
        of our interpretable framework.
        It not only groups anomalies by their statistical properties but also successfully separates them
        into astrophysically coherent classes of anomalies as discussed in Section~\ref{results: clustering}.
    \end{appendix}
\end{document}